\documentclass[conference]{IEEEtran}
\IEEEoverridecommandlockouts
\usepackage{cite}
\usepackage{amsmath,amssymb,amsfonts}
\usepackage{algorithmic}
\usepackage{graphicx}
\usepackage{textcomp}
\usepackage{xcolor}
\usepackage{hyperref}
\usepackage{subfigure}
\def\BibTeX{{\rm B\kern-.05em{\sc i\kern-.025em b}\kern-.08em
    T\kern-.1667em\lower.7ex\hbox{E}\kern-.125emX}}
\begin{document}

\title{A Study on Cognitive Effects of Canvas Size for Augmenting Drawing Skill}

\author{\IEEEauthorblockN{Jize Wang\IEEEauthorrefmark{2}, 
Kazuhisa Nakano\IEEEauthorrefmark{2}, 
Daiyannan Chen\IEEEauthorrefmark{2}, 
Zhengyu Huang\IEEEauthorrefmark{3}, 
\\Tsukasa Fukusato\IEEEauthorrefmark{4}, 
Kazunori Miyata\IEEEauthorrefmark{2}, 
\IEEEauthorrefmark{1}Haoran Xie\IEEEauthorrefmark{2}}

\IEEEauthorblockA{\IEEEauthorrefmark{2}Japan Advanced Institute of Science and Technology, \IEEEauthorrefmark{4}Waseda University}
\IEEEauthorblockA{\IEEEauthorrefmark{3}The Hong Kong University of Science and Technology (Guangzhou)}}

\maketitle

\newcommand\blfootnote[1]{%
  \begingroup
  \renewcommand\thefootnote{}\footnote{#1}%
  \addtocounter{footnote}{-1}%
  \endgroup
}
\blfootnote{\IEEEauthorrefmark{1}Corresponding author (xie@jaist.ac.jp).} 

\begin{abstract}
In recent years, the field of generative artificial intelligence (AI), particularly in the domain of image generation, has exerted a profound influence on society. Despite the capability of AI to produce images of high quality, the augmentation of users' drawing abilities through the provision of drawing support systems emerges as a challenging issue. In this study, we propose that a cognitive factor—specifically, the size of the canvas—may exert a considerable influence on the outcomes of imitative drawing sketches when utilizing reference images. To investigate this hypothesis, a web-based drawing interface was utilized, designed specifically to evaluate the effect of the canvas size's proportionality to the reference image on the fidelity of the drawings produced. The findings from our research lend credence to the hypothesis that a drawing interface, featuring a canvas whose dimensions closely match those of the reference image, markedly improves the precision of user-generated sketches.
\end{abstract}

\begin{IEEEkeywords}
Augmenting Drawing Skill, Reproduction Drawing, Canvas Size, User Interface
\end{IEEEkeywords}

\section{Introduction}
In recent years, the field of generative artificial intelligence (AI), for example, generating high-quality images from users' sketches, has exerted a profound influence on society~\cite{voynov2023sketch, peng2023sketch}. 
These techniques require a dataset with pairwise images and sketches in advance, which is one of the major difficulties in AI research. 
One approach is to hire professional artists to draw a lot of sketches. According to a survey by the Agency for Cultural Affairs (Government of Japan), the population of sculptors, painters, and craftsmen has been between 260,000 and 400,000 from 1989 to 2015 in Japan~\cite{Collection}. 
This approach can guarantee the quality of the sketches, but places a burden on the artists (e.g., one professional artist draws more than 1000 sketches).
On the other hand, if sketches drawn by general users, including novice users (more than 100 million people), are available, it may be possible to reduce time/labor costs and a monetary expense, and easily build sketch2image applications.

\begin{figure}[t]
    \centering
    \subtable[Small size]{\includegraphics[width=0.8\linewidth]{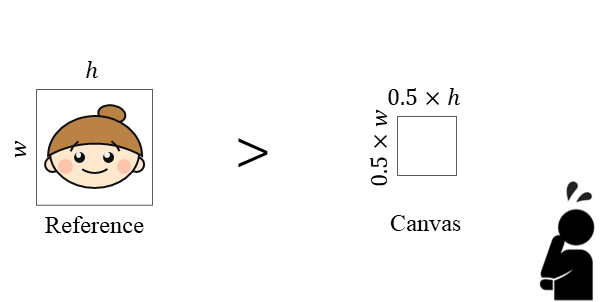}}
    \subtable[Standard size]{\includegraphics[width=0.8\linewidth]{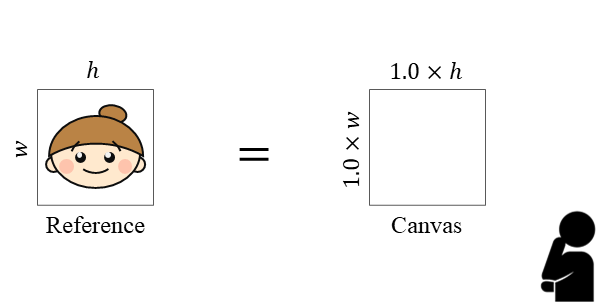}}
    \subtable[Large size]{\includegraphics[width=0.8\linewidth]{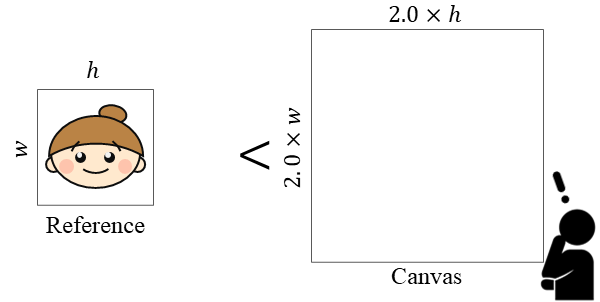}}
    \caption{Comparing the canvas size we prepared in the drawing experiment.}
    \label{fig:size}
\end{figure}

In this paper, as a first step, we focus on the canvas size which is one of the factors that affect the visual quality of users' sketches, and investigate which size is easier for general users to draw (see \autoref{fig:size}). 
This study aims to clarify how the size of the drawing canvas contributes to the improvement of visual details in the drawing process for general users. Note that visual detail refers to the degree to which a reference image can be accurately and faithfully reproduced in the reproduction drawing. 
From the results of this experiment, we can easily provide general users with an appropriate canvas size for creating high-quality pairwise data.

\section{Related Works}

\subsection{Drawings with reference images}
Cohen~\cite{Looklittle} investigated the impact of gaze frequency on drawing accuracy based on the fact that artists have a higher ability to accurately analyze visual stimuli than general people. The results showed that a higher gaze frequency leads to less reliance on long-term memory, reducing memory distortion and enabling accurate drawings. 
Recently, Onishi et al.~\cite{Extraction} conducted a demonstration experiment for feature extraction for better portrait drawing using an eye-tracking device. Note that a third person evaluated the quality of the submitted works after their experiment. The results confirmed that those who were good at drawing made repeated and sufficient observations compared to those who were not good at drawing. This conclusion affirms the conclusion of the previous research~\cite{Looklittle}. Subsequently, a series of empirical studies~\cite{Outlining,Faceprocessing} were conducted on artists and ordinary people, and it was confirmed in other related studies that drawing behavior dependent on short-term memory by frequently aborting reference images improves drawing accuracy. They also concluded that the improvement of drawing accuracy by drawing behavior dependent on short-term memory is common not only in drawing tasks for faces but also in drawing tasks for other objects, such as houses. These studies suggest that visual stimuli have a significant impact on visual detail. 
Inspired by these works, we aim to clarify whether the drawing environment affects the improvement of visual details in the users' drawings.

\subsection{Drawing Support Systems}
Various user interfaces have been proposed to support the improvement of users' drawing skills; Shadow Draw showed that even amateurs can draw high-quality pictures by presenting users with shadow-like drawing guidance in the drawing area~\cite{lee2011shadow}. Kanayama et al. proposed a system to support practice copying by using illustration recommendations of reference images and a variable grid for drawing~\cite{kanayama2023idi}. Recently, with the development of deep learning techniques, several drawing support systems using sketch2image generative models~\cite{DualFace} and latent space exploration~\cite{huang2023anifacedrawing} have been proposed. These methods mainly focus on the drawing target, but they do not mention the quality of their dataset with pairwise sketches and images. In addition, to our knowledge, no research has been published that focuses on the drawing canvas itself, which is one of the most important factors that affects the visual quality of users' sketches.
Therefore, we focus on the effect of the drawing canvas.

\begin{figure}[t]
    \centering
    \includegraphics[width=\linewidth]{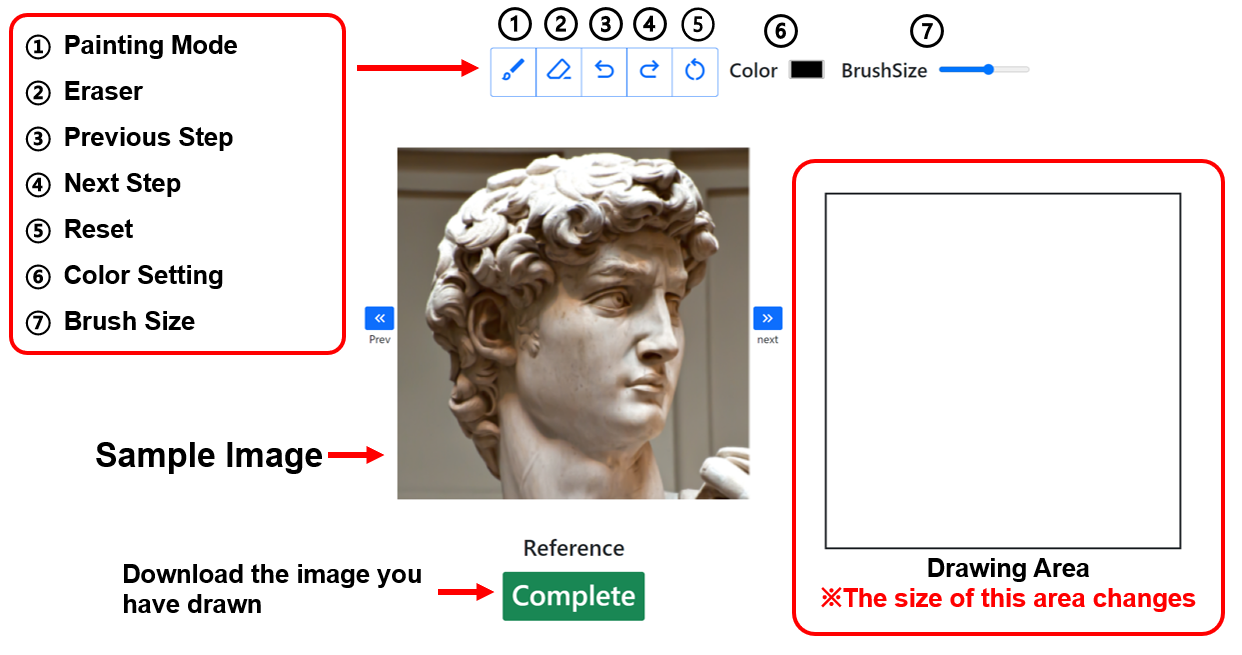}
    \caption{Drawing Interface of the painting system.
    \label{fig:interface}}
\end{figure}


\section{Experiment Design}
We developed a web-based painting system for drawing experiments (see \autoref{fig:interface}). The painting system is based on the size of the reference image ($=$~$500$px~$\times$~$500$px) and provides the following three different canvas sizes, as shown in \autoref{fig:size}.

\begin{figure}[t]
    \centering
    \includegraphics[width=\linewidth]{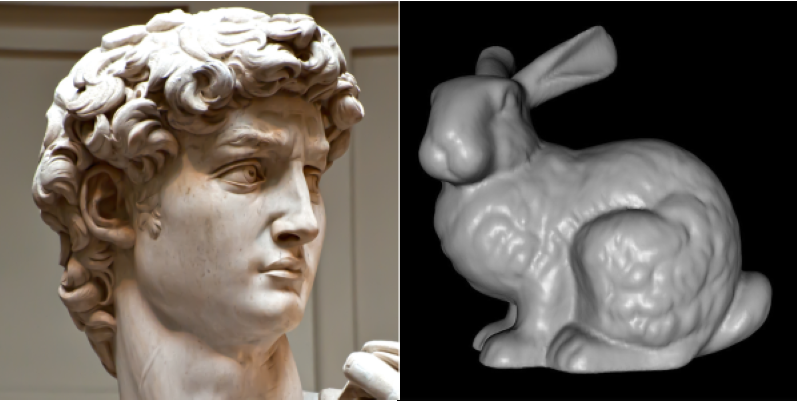}
    \caption{Reference Images: (left)~David and (right)~Stanford Bunny. These sizes are $500$px~$\times$~$500$px. Data from~\cite{Sculpting12}.
    \label{davidandbunny}}
    \label{fig:reference}
\end{figure}

\begin{figure*}[t]
    \centering
    \subtable[David.]{\includegraphics[width=0.9\linewidth]{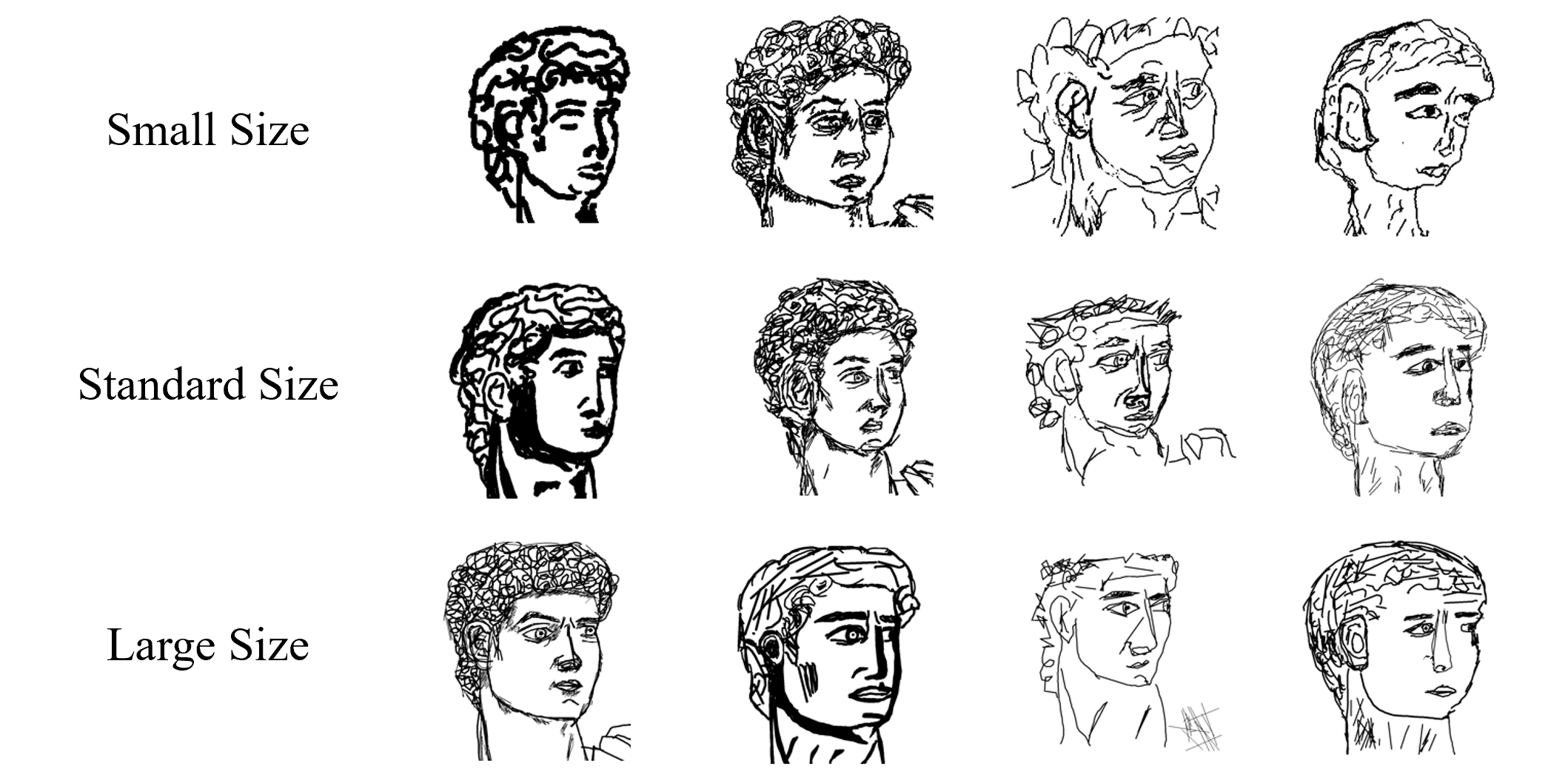}}
    \subtable[Stanford Bunny.]{\includegraphics[width=0.9\linewidth]{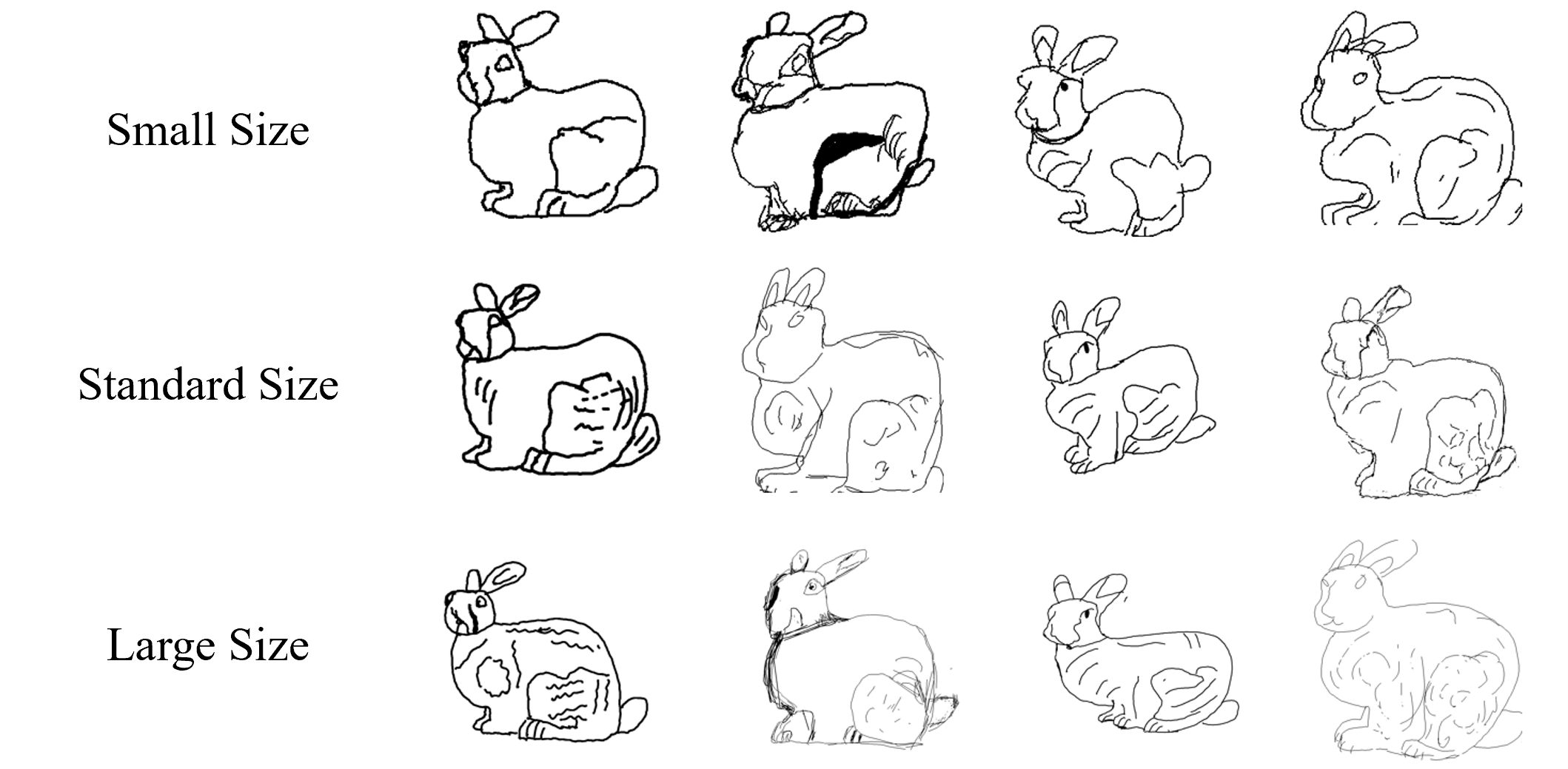}}
    \caption{Examples of reproduction results drawn by participants.}
    \label{fig:drawnresult}
\end{figure*}

\begin{figure*}[t]
    \centering
    \subtable[I would like to use this canvas size frequently.]{\includegraphics[width=0.48\linewidth]{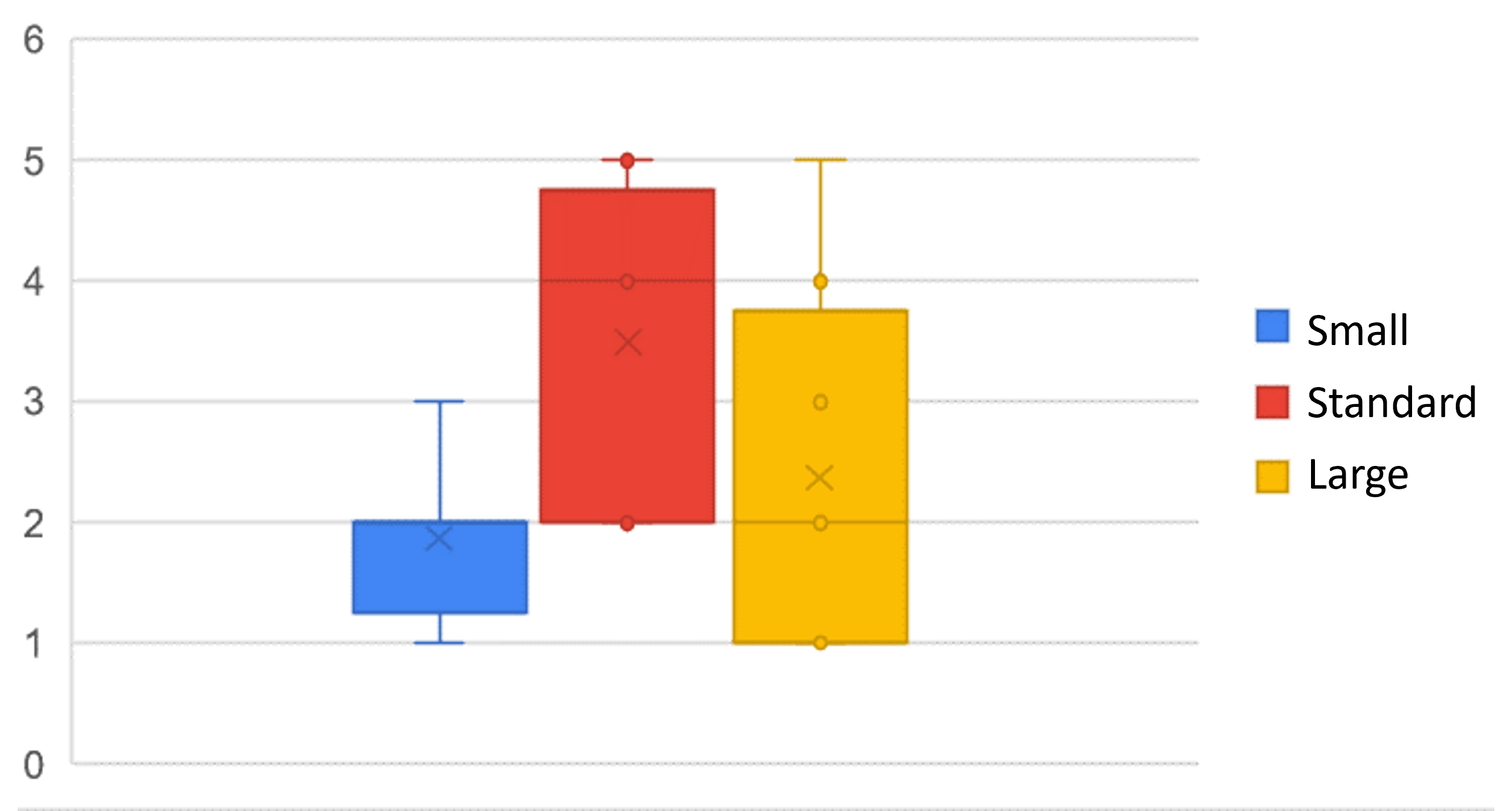}}
    \subtable[I found this canvas size was easy to use for reproducing.]{\includegraphics[width=0.48\linewidth]{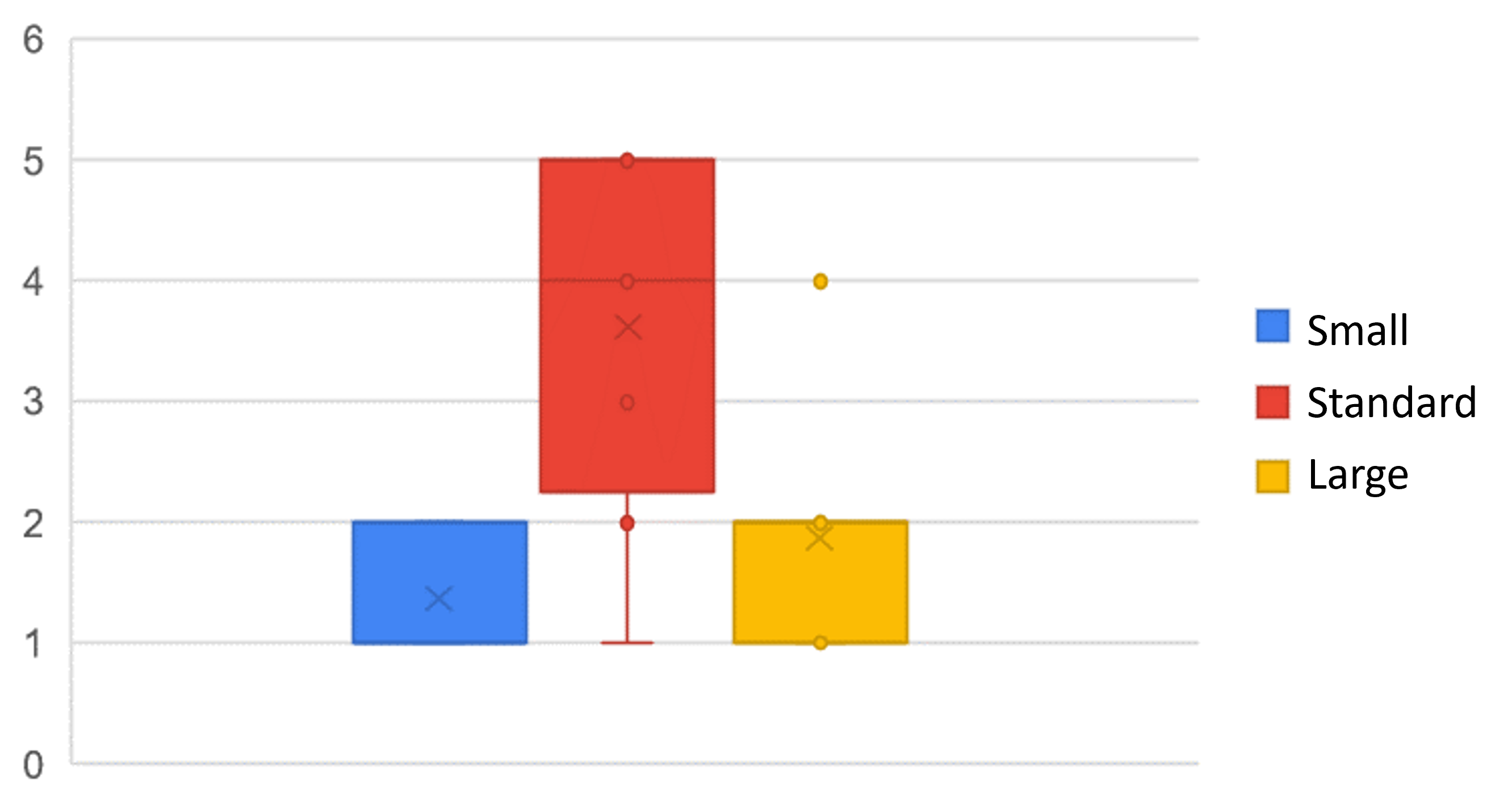}}
    \subtable[I found that most people could reproduce with this canvas size without any hesitation.]{\includegraphics[width=0.48\linewidth]{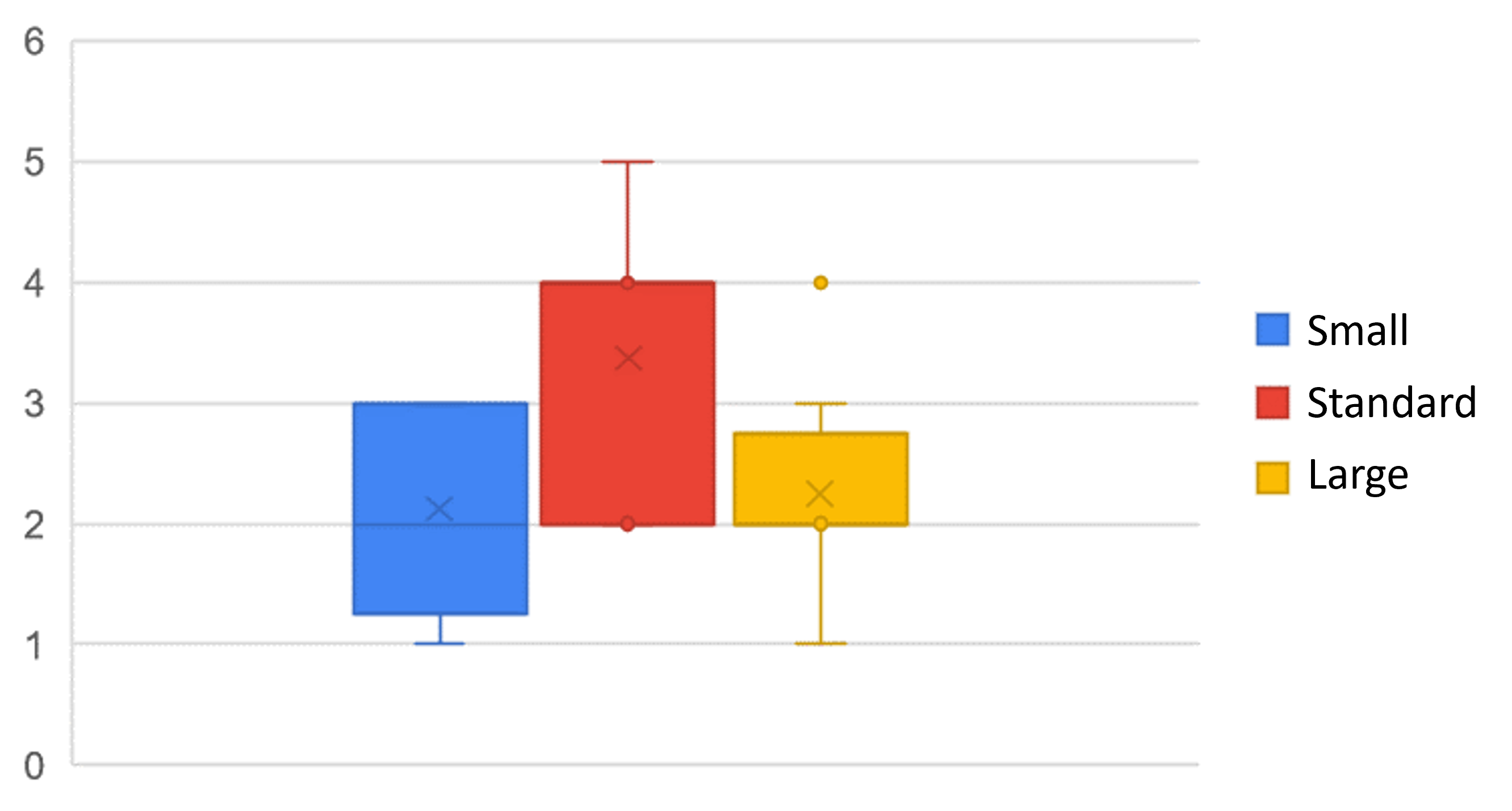}}
    \subtable[I felt comfortable in drawing and reproducing without feeling uneasy.]
    {\includegraphics[width=0.48\linewidth]{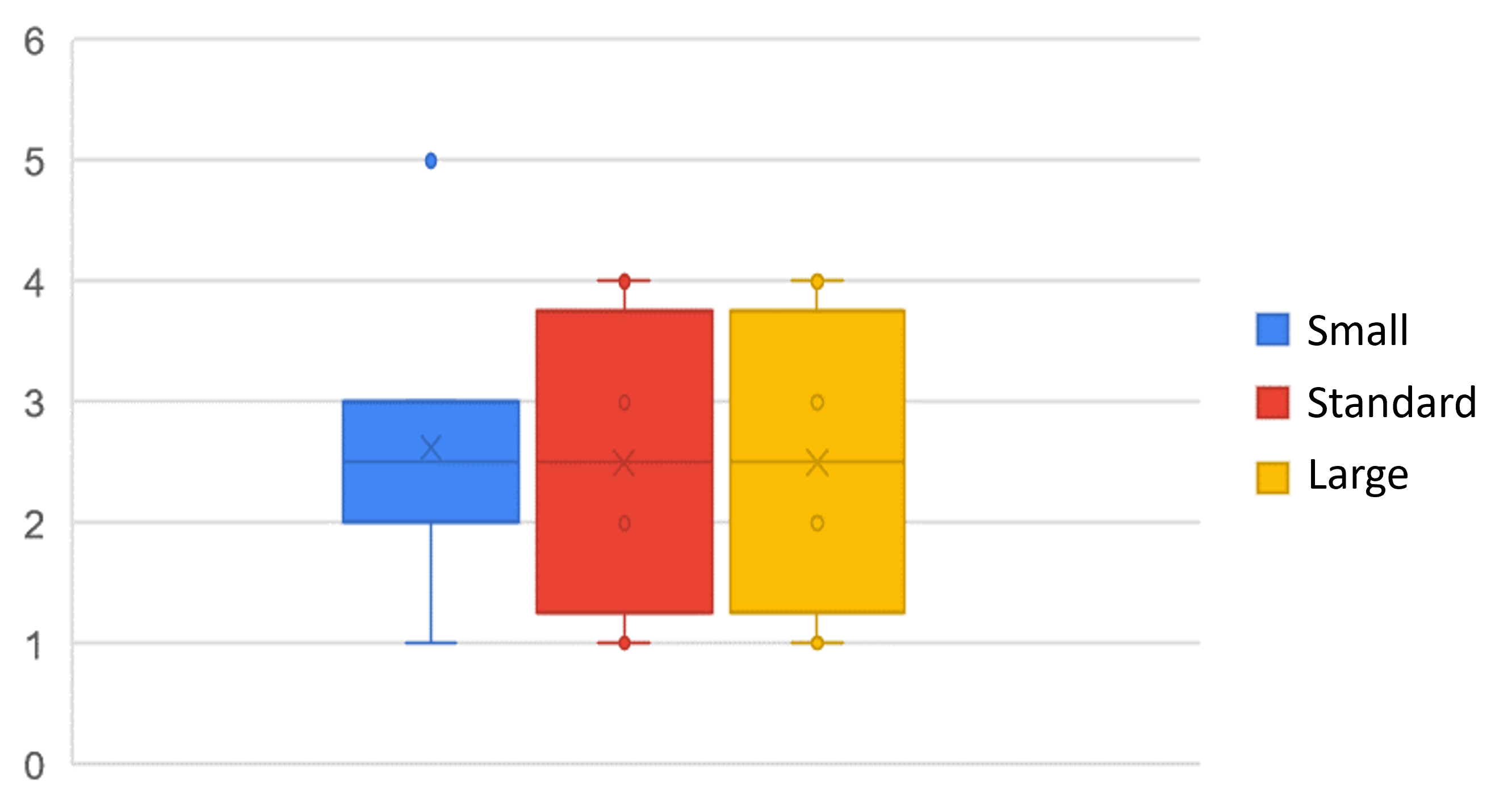}}
    \caption{Questionnaire results of positive question groups.}
    \label{fig:positive}
\end{figure*}

\vspace{2mm}
\begin{enumerate}
    \item A canvas of the same size as reference image, named ``standard size'' ($=$~$500$px~$\times$~$500$px)
    \item A canvas smaller than reference image, named ``small size'' ($=$~$250$px~$\times$~$250$px)
    \item A canvas larger than reference image, named ``large size'' ($=$~$1000$px~$\times$~$1000$px)
    \label{fig:canvassize}
\end{enumerate}
\vspace{2mm}


We conducted experiments to compare the differences between these sizes. We recruited 8 participants (2 males and 6 females in their 20s). Each participant was asked to fill out a pre-questionnaire aimed to investigate his/her participants' drawing skills and prior knowledge. 

%
In the drawing task, we asked the participants to draw and reproduce a given reference image using the developed painting interface (\autoref{fig:interface}) on a Windows laptop (Microsoft Surface 9) with a stylus pen. 
Note that since the reference images should be three-dimensional objects that easily show differences in the participants' drawing ability, we empirically chose the images of David and the Stanford Bunny, as shown in \autoref{fig:reference}.
Regarding time constraints, in the preliminary experiment, we initially set a time limit for each drawing task, but this restriction led to undesirable results. Therefore, we removed the time constraints to allow participants more freedom in their creative process, which yielded better results. 

After the drawing task, each participant filled out post-questionnaires about the effects of the canvas size. The post-questionnaire is based on the System Usability Scale~(SUS), but we selected several items necessary for this experiment from the original SUS questionnaire and evaluated them. The main reason is that our purpose is not to evaluate the painting system itself, but to investigate the participants' awareness of each canvas size. 

%
The total experiment took approximately 60 minutes per participant.


\begin{figure}[t]
    \centering
    \subtable[I felt that this canvas made reproducing more complicated than necessary.]{\includegraphics[width=0.99\linewidth]{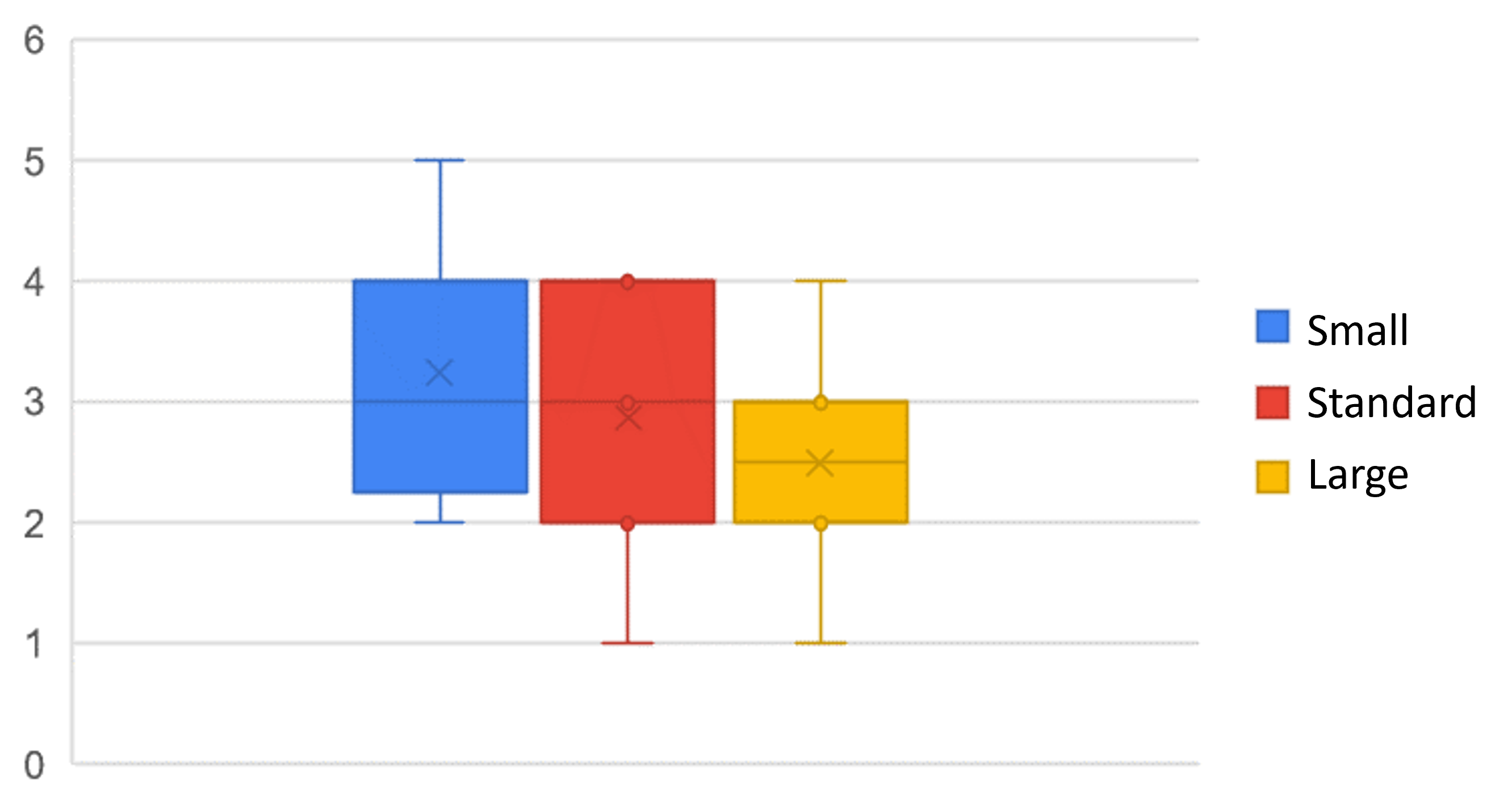}}
    \subtable[I found this canvas very difficult to draw on.]{\includegraphics[width=0.99\linewidth]{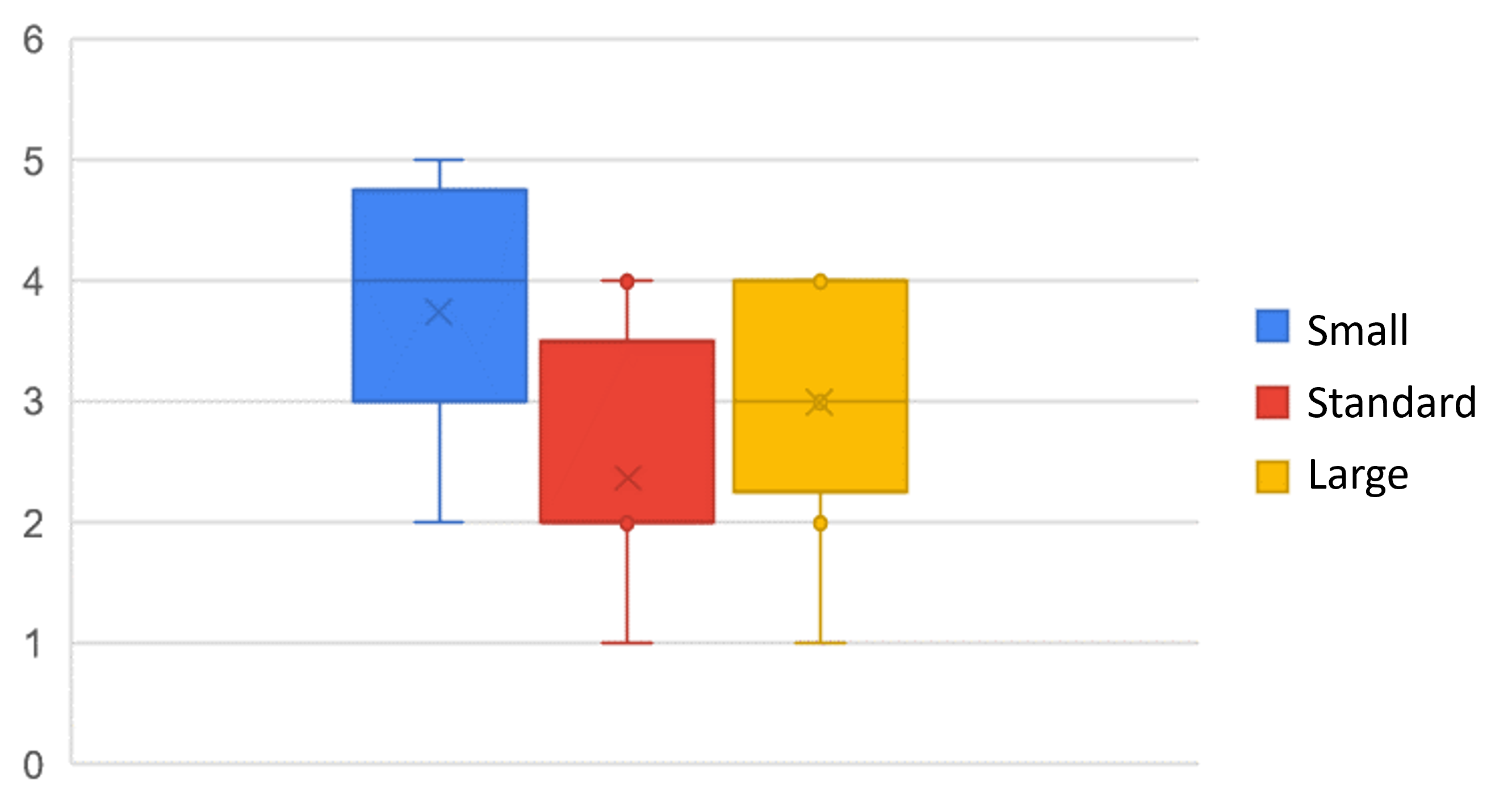}}
    \caption{Questionnaire results of negative question groups.}
    \label{fig:negative}
\end{figure}

\section{Results}
In the experiment, about the design experiences with arts education, the majority of participants~($=$ 87.5\%) have not received an art-related education. \autoref{fig:drawnresult} shows examples of reproduction drawing results drawn by participants.


\subsection{Questionnaire Results}
The results of the reproduction drawing experiment show that the standard size is the optimal canvas size for users. The average values for the positive items in the questionnaire showed that the standard size had a higher value compared to other items (see \autoref{fig:positive}). In particular, the average value of the item ``\textit{easy to use for reproducing}'' is higher than that of the other sizes. On the other hand, the results for the question ``\textit{I felt comfortable in drawing and reproducing without feeling uneasy},'' were uniform for all items.

The average value for the negative questions was relatively high for the small size, indicating that participants had more difficulties with the smaller canvas size, as shown in \autoref{fig:negative}. However, the large size had the lowest average value for the item ``\textit{Did the canvas size make the reproduction drawing unnecessarily complex~?},'' suggesting that participants found the larger canvas size to be less complex and more comfortable for reproduction drawing.

Based on the questionnaire results, it is clear that the standard size is the preferred canvas size for the participants. However, there was no clear difference in the evaluation of confidence in being able to reproduce without difficulty, or in the item of making the reproduction more complicated than necessary.

\subsection{Free Comments of Participants}
The results of the free comments revealed that approximately 70\% of the participants thought that the standard size is helpful for them to faithfully draw and reproduce the reference image, and the remaining 30\% of the participants preferred the large size. Specific comments are shown below.

\begin{itemize}
    \item \textit{The canvas size is close to the reference image, so it is easier to determine where to draw.}
    \item \textit{It is easier to understand the placement of lines because the size of the object being drawn is the same.}
    \item \textit{When the canvas is large, it is easy to paint with a high degree of freedom, but at the same time, it is difficult to know how much detail to express.}
    \item \textit{The larger size can be drawn in many spaces.}
    \item \textit{When drawing at a small size, the curves are not drawn perfectly, and the result is more poorly drawn than expected.}
\end{itemize}

There were many responses regarding the blank space in the drawing area and comparing the canvas size with the reference image. Especially, the participants' responses, such as ``\textit{The canvas size is close to the reference image, so it is easier to determine where to draw}'' and ``\textit{It is easier to understand the placement of lines because the size of the object being drawn is the same},'' suggest that participants expect the canvas size to match the space of the reference image in order to trace it more easily when engaging in faithful reproduction drawing. 
On the other hand, one possible reason why some participants preferred the large size is that the large canvas makes it easier to draw the participants' own interpretation.

\section{Conclusion}
In this study, we conducted a comparison experiment on the ease of drawing for three different sizes (i.e., small size, standard size, and large size) based on the size of the reference image for augmenting drawing skills. The results of the questionnaire showed that the standard size was relatively easy to draw and had a high level of faithfulness based on the subjective evaluation of the participants. 

In addition, we found the participants were able to draw local details on the large canvas, but at the same time, they often added their own interpretation. 
This indicates that the large size may have a negative impact on the faithfulness of the reproduction drawing, which is the opposite of what has been found in previous research.


This experiment's results suggest that the standard size, which was found to be the most preferred and easy to use for participants, may also be the optimal size for ensuring faithfulness in the reproduction process. However, further research is needed to confirm these results and to explore the relationship between the size of the reference image and the faithfulness of the reproduction drawing. 
We will also explore ways to interactively change the reference image itself instead of changing the canvas size. 
The current experiment did not address the evaluation of the quality of user-designed sketches and the impact of participant gaze. In future studies, the relationship between the visual quality of artwork and the difference in the gaze movements caused by the canvas size will be investigated.

\section*{Acknowledgment}
This research has been supported by JAIST research fund, and the Kayamori Foundation of Informational Science  Advancement. We thank the anonymous reviewers for their insightful comments.

\bibliographystyle{IEEEtran}
\bibliography{IEEEabrv, refs}
\end{document}